\documentclass[a4paper,11pt]{article}\usepackage{jcappub}\def\PRDstyle#1{}\def\JCAPstyle#1{#1}\let\Abstract\abstract
\usepackage[utf8]{inputenc}
\usepackage[T1]{fontenc}
\usepackage{cmap}
\usepackage{rotating}
\usepackage{float}

\DeclareMathAlphabet{\pazocal}{OMS}{zplm}{m}{n}

\def\imo{i}

\def\K{{\cal K}}

\begin{document}

\title{Quasinormal modes and grey-body factors of regular black holes with a scalar hair from the Effective Field Theory}

\JCAPstyle{
\author[1]{R.~A.~Konoplya,}\emailAdd{roman.konoplya@gmail.com}
\affiliation[1]{Research Centre for Theoretical Physics and Astrophysics, \\ Institute of Physics, Silesian University in Opava, \\ Bezručovo náměstí 13, CZ-74601 Opava, Czech Republic}
\arxivnumber{2305.09187}
}

\Abstract{
The Effective Field Theory (EFT) of perturbations on an arbitrary background geometry with a timelike scalar profile has been recently constructed in the context of scalar-tensor theories. Unlike General Relativity, the regular Hayward metric is realized as an exact background metric in the Effective Field Theory with timelike scalar profile without resorting to special matter field, such as nonlinear electrodynamics. The fundamental quasinormal mode for axial graviational perturbations of this black hole has been considered recently with the help of various methods. Here we make a further step in this direction and find that, unlike the fundamental mode, a few first overtones deviate from their Schwarzschild limit at a much higher rate. This outburst of overtones occurs because the overtones are extremely sensitive to the least change of the near- horizon geometry. The analytical formula for quasinormal modes is obtained in the eikonal regime. In addition, we calculated grey-body factors and showed that the regular Hayward black hole with a scalar hair has a smaller grey-body factor than the Schwarzschild one.  Integration of the wave-like equation in the time-domain shows that the power-law tails, following the ring-down phase,  are indistinguishable from the Schwarzschild ones at late times.
}

\maketitle

\section{Introduction}\label{Introduction}

The Effective Field Theory (EFT) offers a model-independent framework for studying the dynamics of perturbations on a particular spacetime. The fundamental components of the EFT include the configuration of matter fields, the symmetry breaking mechanism, and the background spacetime. The EFT of scalar-tensor theories on Minkowski and de Sitter spacetimes, known as ghost condensation, was formulated long time ago in \cite{Arkani-Hamed:2003pdi,Arkani-Hamed:2003juy}, under the assumption that a scalar field's timelike gradient spontaneously breaks time diffeomorphism and that the EFT remains invariant under spatial diffeomorphism, time translation, and time reflection, subject to the shift and reflection of the scalar field. Recent work \cite{Mukohyama:2022enj} generalized the EFT action with a timelike scalar profile to arbitrary background geometries. Then, a dictionary was developed to relate this EFT to concrete covariant theories such as the Horndeski's theory~\cite{Mukohyama:2022enj} and shift-symmetric quadratic higher-order scalar-tensor (HOST) theories~\cite{Mukohyama:2022skk}. As a result of the scalar field's timelike nature, the EFT of \cite{Mukohyama:2022enj} can describe both cosmological and black hole scales in a single framework, making it possible to extract some information about scalar-field dark energy from astrophysical black hole observations. This feature induces additional interest to the EFT, making it appealing to consider the EFT of black hole perturbations with a timelike scalar profile. It is worth noting that an EFT of black hole perturbations with a spacelike scalar profile on a static and spherically symmetric background was developed in \cite{Franciolini:2018uyq} (see also \cite{Hui:2021cpm} for the formulation of the EFT on a slowly rotating black hole).

Despite perturbations and stability of black holes in scalar-tensor theories were studied in quite a few works  \cite{Babichev:2018uiw}-\cite{Takahashi:2021bml},  the above EFT approach allows one to study perturbations and quasinormal modes, of black holes in scalar-tensor theories in a model-independent manner. The first step in this direction was made in \cite{Mukohyama:2023xyf}, where the formalism for the EFT of perturbations on a static and spherically symmetric black hole with a timelike scalar profile was constructed and the two examples of background solutions were considered: the stealthy Schwarzschild black hole and Hayward spacetime with a non-trivial scalar hair. While the first metric has rather trivial spectrum, which can be easily found via re-scaling of the well-known Schwarzschild spacetime, the second solution is remarkable. First of all, in some range of parameters it represents a regular black hole without resorting to any specific types of matter such as the non-linear electrodynamics with a particular source of non-linearity. Secondly, it has a different quasinormal spectrum from either Schwarzschild metric or Hayward solution in the non-linear electrodynamics. The fundamental mode (for the lowest overtone and multipole numbers) for this case were calculated in \cite{Mukohyama:2023xyf} with the help of the Leaver, direct integration and 6th order WKB methods. The corresponding master wave-like equation describing axial gravitational perturbations was earlier deduced in \cite{Mukohyama:2022skk}.

In this work, we will take a further step and investigate not only the fundamental mode but also several first overtones of the EFT of gravitational perturbations of the Hayward spacetime. Although it is commonly believed that the principal contribution to the signal is due to the fundamental mode, recent research \cite{Giesler:2019uxc} (and subsequently examined in \cite{Oshita:2021iyn,Forteza:2021wfq,Oshita:2022pkc}) has demonstrated that in order to simulate the ringdown phase accurately, which is obtained within the precise numerical relativity simulations at the start of the quasinormal ringing and not just at the final stage, the first few overtones must be considered. This discovery also suggested that the actual QN ringing begins earlier than expected. Once the overtones are taken into account, the modelling and linear profile of the ringdown are in full agreement, allowing for the extraction of the black hole's angular mass and momentum. Despite the current LIGO/VIRGO observational data not allowing the detection of the overtones in the gravitational-wave signals \cite{Capano:2021etf,Cotesta:2022pci}, and the higher-overtone contributions requiring nonlinear corrections \cite{Sberna:2021eui,Cheung:2022rbm,Mitman:2022qdl} when analyzing them, there are indications that the overtones are significantly excited in some events and may be detectable by LISA during the early ringdown phase \cite{Oshita:2022yry}.

An important aspect related to the overtones' behavior is related to the near horizon geometry of a black hole. If modifications to Einstein's theory induce noticeable deformations of the black-hole geometry only near the event horizon, the fundamental mode remains largely unaffected. However, even a small change near the event horizon can significantly impact the first few overtones \cite{Konoplya:2022pbc}, providing a means to probe the geometry of the event horizon. As was shown in \cite{Konoplya:2022pbc} overtones are stable against small deformations of spacetime at a distance from the black hole, allowing the event horizon to be distinguished from the surrounding environment. In contrast to echoes, overtones make a much larger energy contribution. This finding gives us further motivation to study overtones of various black holes \cite{Konoplya:2022hll,Konoplya:2022iyn,Konoplya:2023aph}.

Taking into consideration the above motivations we will study here several first overtones for the axial gravitational perturbations following from the Effective Field Theory with timelike scalar profile. In addition we will consider time-domain integration of the above wave-like equation which allow us to see the late time tails following the ring-down phase. We will also study grey-body factors of the gravitational perturbations.

The paper is organized as follows. In Sec. II we will briefly review  perturbations of the Haywrad black hole in the Effective Field Theory. Sec. III is devoted to numerical and semi-analytic methods used for calculations of quasinormal modes. Sec. IV discusses quasinormal modes, both fundamental one and outburst of overtones, as well as time-domain evolution of the perturbation. In Sec. V the grey-body factors are considered. Finally, we summarize the obtained results and mention some open question.

\section{Black Hole Perturbations in the Effective Field Theory}\label{BHPertubrations}

Here we will provide a brief overview of the Effective Field Theory (EFT) for perturbations on an arbitrary background with a timelike scalar profile \cite{Mukohyama:2022enj,Mukohyama:2022skk}. For a similar EFT construction for shift-symmetric scalar-tensor theories, see \cite{Khoury:2022zor}. The main concept behind our EFT, which is similar to the EFT of ghost condensation \cite{Arkani-Hamed:2003pdi} and the EFT of inflation/dark energy \cite{Cheung:2007st,Gubitosi:2012hu}, is that the scalar field's time-dependent background, $\bar{\Phi}$, breaks time diffeomorphism spontaneously, establishing a preferred ($\Phi = const.$) time-slicing. This time-slicing can be defined by the unit normal vector, $n_\mu$, as shown in,
\begin{align}\label{eq:normal_EFT}
n_\mu \equiv
-\frac{\partial_\mu \Phi}{\sqrt{-X}}
\rightarrow - \frac{\delta_\mu^\tau}{\sqrt{-g^{\tau\tau}}} \;,
\end{align}
where $X\equiv g^{\mu\nu}\partial_\mu\Phi\partial_\nu\Phi$ is the scalar field's kinetic term, and $n_\mu n^\mu = -1$. Here, we use $\tau$ as the time coordinate such that $\bar{\Phi}=\bar{\Phi}(\tau)$ and $\delta\Phi\equiv\Phi-\bar{\Phi}=0$ (unitary gauge). The expression on the right-hand side of (\ref{eq:normal_EFT}) refers to the one in the unitary gauge. The EFT's residual symmetry, therefore, is 3d diffeomorphism invariance. Consequently, the EFT we will write down in the unitary gauge may contain a scalar function of, for example, the 4d and 3d curvatures, the extrinsic curvature, the ($\tau\tau$)-component of the inverse metric tensor, and the time coordinate $\tau$.

We will use the Arnowitt-Deser-Misner (ADM) $3+1$ decomposition, in which the metric can be expressed as shown in,
\begin{align}
ds^2 = -N^2 d\tau^2 + h_{ij} (dx^i + N^i d\tau) (dx^j + N^j d\tau).
\end{align}
Here, $N$ is the lapse function, $N^i$ is the shift vector, and $h_{\mu\nu} \equiv g_{\mu\nu} + n_\mu n_\nu$ is the induced metric on a spacelike hypersurface of constant $\tau$. The indices $i$ and $j$ refer to the spatial components, and the spatial indices are raised and lowered by the induced metric $h_{ij}$.

The extrinsic curvature is defined using the induced metric~$h_{\mu\nu}$:
 \begin{align}
 K_{\mu\nu} \equiv h_\mu^\rho \nabla_\rho n_\nu.
 \end{align}
Here $\nabla_\mu$ is the 4d covariant derivative. Then, one can find  the spatial components of $K_{\mu\nu}$ and its trace in terms of the ADM variables in the following form:
\begin{align}
K_{ij} = \frac{1}{2N}\left(\dot{h}_{ij} - D_i N_j - D_j N_i\right) \;, \qquad K = h^{ij} K_{ij} \;,
\end{align}
with a dot being the derivative with respect to $\tau$ and $D_i$ the 3d covariant derivative constructed from the induced metric~$h_{ij}$. In addition, the 3d curvature~${}^{(3)}\!R$ can be found via the induced metric~$h_{ij}$.

The unitary-gauge EFT action we are formulating is invariant, that is, not affected by 3d diffeomorphism. Thus, in addition to the 4d covariant terms such as the 4d Ricci scalar~$\tilde{R}$, the action can rely on any geometrical quantities that are covariant under the 3d diffeomorphism, such as $g^{\tau\tau}$ ($=-1/N^2$), $K_{\mu\nu}$, and ${}^{(3)}!R_{\mu\nu}$. It's worth noting that the symbol~$R$ (without the tilde) denotes the 4d Ricci scalar with the divergence term subtracted. Moreover, the action can depend on $\tau$ explicitly. With all the possibilities above, the unitary-gauge action takes the form
\begin{align}\label{eq:action_uni}
S = \int d^4x \sqrt{-g}~F(\tilde{R}_{\mu\nu\alpha\beta}, K_{\mu\nu}, g^{\tau\tau}, \nabla_\mu, \tau) \;,
\end{align}
where $F$ is a scalar function of those 4d and 3d diffeomorphism covariant quantities, and $\tilde{R}_{\mu\nu\alpha\beta}$ is the 4d Riemann tensor of the metric~$g_{\mu\nu}$.
The 3d curvature tensor can be written in terms of the 4d curvature and the extrinsic curvature by use of the Gauss relation, so that one does not include it explicitly in the action. The action~(\ref{eq:action_uni}) could be applied to any background geometries without assuming a particular symmetry of the background.

Following \cite{Mukohyama:2022enj}, the perturbations are defined as follows:
\begin{align}
\delta g^{\tau\tau} \equiv g^{\tau\tau} - \bar{g}^{\tau\tau}(\tau, \vec{x}) \;, \qquad \delta K^\mu_\nu \equiv K^\mu_\nu - \bar{K}^\mu_\nu(\tau, \vec{x}) \;, \qquad \delta {}^{(3)}\!R^\mu_\nu \equiv {}^{(3)}\!R^\mu_\nu -  {}^{(3)}\!\bar{R}^\mu_\nu(\tau, \vec{x}) \;, \label{eq:pert}
\end{align}
where a bar denotes the background value. Practically, the EFT action is written as a polynomial of these perturbation variables as well as their derivatives. Each EFT coefficient can have an explicit dependence on both $\tau$ and $\vec{x}$ due to the spacetime dependence of the background quantities~$\bar{g}^{\tau\tau}$, $\bar{K}^\mu_\nu$, and ${}^{(3)}\!\bar{R}^\mu_\nu$. Note that such a dependence on $\vec{x}$ is not compatible with the 3d diffeomorphism invariance in general. Therefore, in order for the EFT action to respect the 3d diffeomorphism invariance,  a set of consistency relations on the EFT coefficients was imposed. The technically difficult formalism for writing-down the linearized perturbation equations and reduction them to the wave-like form was developed in  \cite{Mukohyama:2023xyf} and we refer a reader to this work for details.

The Hayward metric~\cite{Hayward:2005gi} which is the background solution in the above Effective Field Theory has the form:
\begin{equation}
g_{tt} = -g_{rr}^{-1} =1-\frac{\mu r^2}{r^3+\sigma^3}.
\end{equation}
For $\sigma > 0$, the Hayward metric corresponds to a regular black hole, but not at $\sigma < 0$, as there exists a curvature singularity at $r = - \sigma$. One
could regard $\sigma$ as just a phenomenological parameter that controls the deviation from the Schwarzschild metric. The perturbation equations, however, depend not upon the event horizon of the above metric function, but on a new function, corresponding to axial gravitational perturbations built from EFT. From here and on, we will use the resultant wave like equation for the axial gravitational perturbations in the EFT deduced in \cite{Mukohyama:2023xyf}:
\begin{align}\label{eq:Q_master}
\frac{d^2}{dr_*^2}\Psi(r_*) + (\omega^2 - V_{\rm eff})\Psi(r_*) = 0 \;.
\end{align}

The function~$F\equiv dr/dr_*$ is
    \begin{align}\label{eq:drdrs-Hayward}
    F(r)=\frac{r^4 - \mu (r^3 + \sigma^3)}{r (r^3 + \sigma^3)}\;.
    \end{align}
The position of the odd-mode horizon~$r_g\,(>0)$ is given by $F(r_g)=0$, or equivalently,
    \begin{align}
    r_g^4-\mu(r_g^3+\sigma^3)=0\;,
    \label{eq:r_g-Hayward}
    \end{align}
which has a single positive solution so long as $\mu$ and $\sigma$ are positive.
The effective potential is
    \begin{align}
    V_{\rm eff}(r)
    =\left[1-\frac{\mu(r^3+\sigma^3)}{r^4}\right]
    \left\{\frac{\ell(\ell+1)r^4}{(r^3+\sigma^3)^2}-\frac{3\left[4\mu r^9+2\sigma^3r^6(8r-\mu)+\sigma^6r^3(r-7\mu)-\mu\sigma^9\right]}{4(r^3+\sigma^3)^4}\right\}\;.
    \label{eq:Hayward_potential}
    \end{align}
Instead of choosing $r_g$ and $\sigma$ as independent parameters and fixing $\mu$ we fix the radius of the event horizon $r_g =1$ and change $\sigma$, so that
    \begin{align}
    \mu=\frac{1}{1+\sigma^3}\;.
    \end{align}
The above effective potential is positive definite and has single maximum (see fig. \ref{fig2}).

%\begin{small}
\begin{sidewaystable}
%\begin{adjustbox}{angle=90}
    \centering
%\begin{sideways}
\caption{First three quasinormal modes for $\ell=2$ and various values of $\sigma$ calculated by the 7th order WKB method with Pade apprximants and accurate Leaver method. Notice, that the units used here, $r_g =1$, is different from those of \cite{Mukohyama:2023xyf}, which are $\mu =1$.}
\label{Table1}
\begin{tabular}{|c|c|c|c|c|c|c|}
  \hline
  \hline
   $\sigma$ & $n=0$ (WKB) & $n=0$ (Leaver) & $n=1$ (WKB)  & $n=1$(Leaver)  &$n=2$ (WKB)  & $n=2$ (Leaver) \\
  \hline
    \hline
  -0.4 & $0.72808 - 0.16273 i$ & $0.728052 - 0.162829 i$ & $0.68425 - 0.49887 i$ & $0.684225 - 0.499515 i$  & $0.60706 - 0.86454 i$ & $0.609478 - 0.866055 i$\\
  -0.3 & $0.74016 - 0.17181 i$ & $0.740213 - 0.171869 i$ & $0.69004 - 0.52775 i$ & $0.690847 - 0.528492 i$  & $0.60345 - 0.91945 i$ & $0.607254 - 0.920381 i$ \\
  -0.2 & $0.74531 - 0.17613 i$ & $0.745369 - 0.176181 i$ & $0.69166 - 0.54158 i$ & $0.692822 - 0.542267 i$  & $0.59894 - 0.94506 i$ & $0.603861 - 0.946149 i$\\
  -0.1 & $0.74704 - 0.17766 i$ & $0.747103 - 0.177709 i$ & $0.69213 - 0.54639 i$ & $0.693354 - 0.547142 i$  & $0.59722 - 0.95403 i$ & $0.602335 - 0.955268 i$\\
  0    & $0.74729 - 0.17787 i$ & $0.747343 - 0.177925 i$ & $0.69220 - 0.54707 i$ & $0.693422 - 0.547830 i$  & $0.59697 - 0.95529 i$ & $0.602107 - 0.956554 i$\\
  0.1  & $0.74752 - 0.17809 i$ & $0.747582 - 0.178139 i$ & $0.69226 - 0.54774 i$ & $0.693488 - 0.548515 i$  & $0.59672 - 0.95655 I$ & $0.601876 - 0.957836 i$\\
  0.2  & $0.74915 - 0.17957 i$ & $0.749208 - 0.179624 i$ & $0.69267 - 0.55235 i$ & $0.693900 - 0.553250 i$  & $0.59500 - 0.96519 i$ & $0.600195 - 0.966697 i$ \\
  0.3  & $0.75316 - 0.18340 i$ & $0.753224 - 0.183484 i$ & $0.69362 - 0.56417 i$ & $0.694606 - 0.565556 i$  & $0.59000 - 0.98734 i$ & $0.595137 - 0.989760 i$ \\
  0.4  & $0.75956 - 0.19007 i$ & $0.759561 - 0.190294 i$ & $0.69447 - 0.58455 i$ & $0.694590 - 0.587283 i$  & $0.57925 - 1.02328 i$ & $0.583917 - 1.030721 i$ \\
  0.5  & $0.76674 - 0.19910 i$ & $0.766538 - 0.199606 i$ & $0.69231 - 0.60992 i$ & $0.691925 - 0.617129 i$  & $0.55996 - 1.05669 i$ & $0.564135 - 1.088089 i$\\
  0.6  & $0.77158 - 0.20896 i$ & $0.771162 - 0.209718 i$ & $0.68398 - 0.63594 i$ & $0.684964 - 0.649982 i$  & $0.51059 - 1.06490 i$ & $0.537282 - 1.154489 i$\\
  0.7  & $0.77061 - 0.21740 i$ & $0.770339 - 0.218185 i$ & $0.67149 - 0.65870 i$ & $0.673597 - 0.678287 i$  & $0.46298 - 1.06576 i$ & $0.511096 - 1.218028 i$\\
  0.8  & $0.76151 - 0.22231 i$ & $0.762131 - 0.222897 i$ & $0.65316 - 0.67322 i$ & $0.659015 - 0.695039 i$  & $0.45630 - 1.06699 i$ & $0.494865 - 1.264176 i$\\
  0.9  & $0.74425 - 0.22337 i$ & $0.746259 - 0.222918 i$ & $0.62929 - 0.67911 i$ & $0.642316 - 0.696476 i$  & $0.45957 - 1.09357 i$ & $0.489861 - 1.280763 i$\\
  1.0  & $0.72009 - 0.22301 i$ & $0.723789 - 0.218567 i$ & $0.60258 - 0.68871 i$ & $0.623701 - 0.682879 i$  & $0.44887 - 1.16460 i$ & $0.489066 - 1.265066 i$\\
  \hline
   \hline
\end{tabular}
%\end{adjustbox}
%\end{sideways}
  \end{sidewaystable}
%\end{small}
  \begin{figure}
\centerline{\resizebox{\linewidth}{!}{\includegraphics*{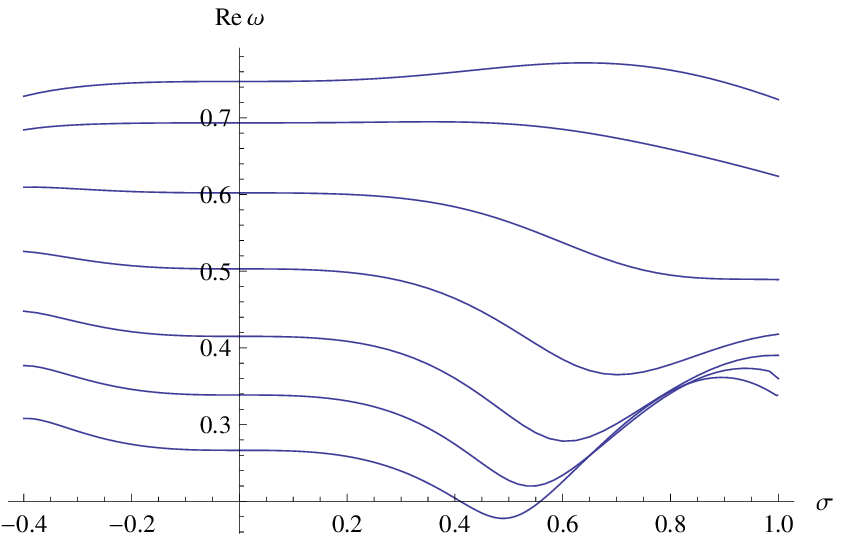}\includegraphics*{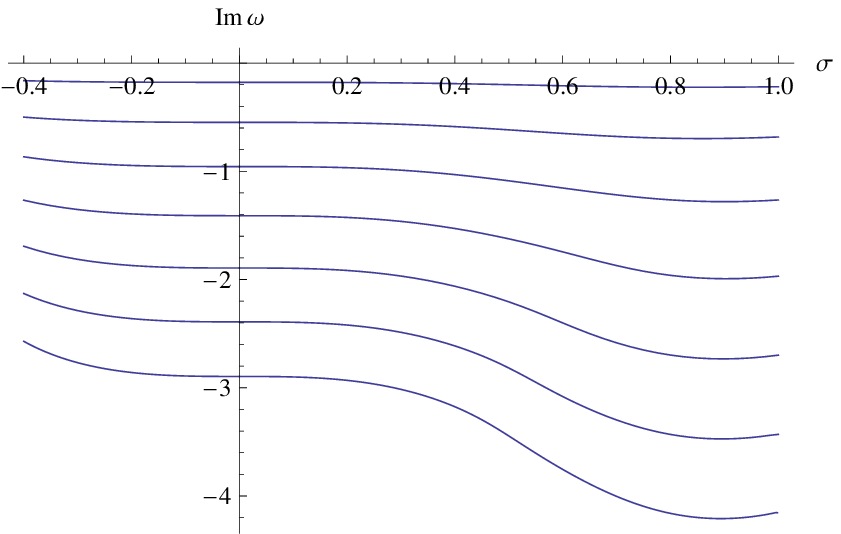}}}
\caption{$Re \omega$ (left) and $Im \omega$ (right) as a function of $\sigma$ (in units $r_{g}=1$) for the fundamental mode and first six overtones from top to bottom.}\label{fig1}
\end{figure}
\begin{figure}
\centerline{\resizebox{0.7 \linewidth}{!}{\includegraphics*{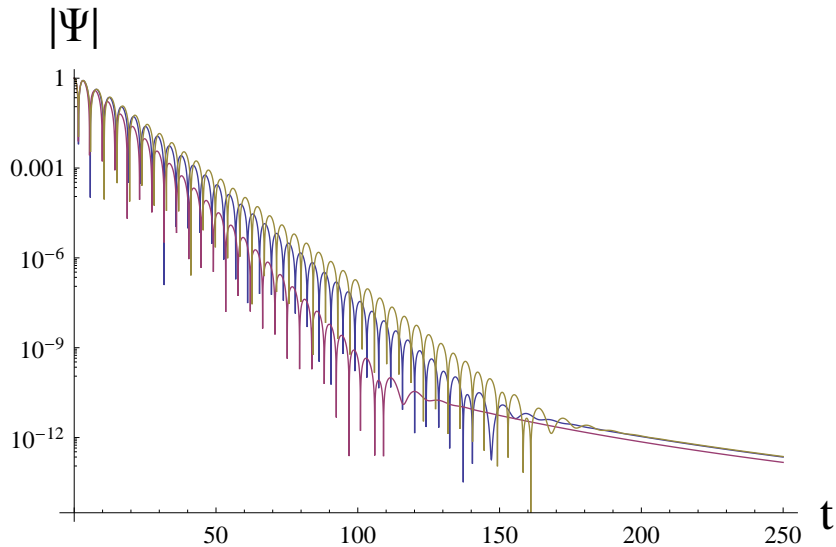}}}
\caption{Semi-logarithmic time domain profile for $\ell=2$, $\sigma =0$ (blue, middle), $\sigma =-0.4$ (green, top) and $\sigma =1$ (red, bottom).}\label{fig1}
\end{figure}\label{fig3}

\section{Methods used for calculation of quasinormal modes}\label{Methods}

Here we will briefly review the three methods used for the spectral analysis of black hole perturbations: WKB method, Frobenius method and time-domain integration.

\subsection{WKB method}
In the frequency domain we will use the semi-analytic WKB approach applied by Will and Schutz \cite{Schutz:1985km} for finding quasinormal modes. The Will-Schutz formula was extended to higher orders in \cite{Iyer:1986np,Konoplya:2003ii,Matyjasek:2017psv} and made even more accurate when using the Padé approximants \cite{Matyjasek:2017psv,Hatsuda:2019eoj}.
The general WKB formula has the form \cite{Konoplya:2019hlu},
\begin{eqnarray}
\omega^2&=&V_0+A_2(\K^2)+A_4(\K^2)+A_6(\K^2)+\ldots \\\nonumber
&-& \imo \K\sqrt{-2V_2}\left(1+A_3(\K^2)+A_5(\K^2)+A_7(\K^2)+\ldots\right),
\end{eqnarray}
where $\K=n+1/2$ is half-integer. The corrections $A_k(\K^2)$ of the order $k$ to the eikonal formula are polynomials of $\K^2$ with rational coefficients and depend on the values of higher derivatives of the potential $V(r)$ in its maximum. In order to increase the accuracy of the WKB formula, we will follow the procedure of Matyjasek and Opala \cite{Matyjasek:2017psv} and use the Padé approximants. Here we will use the sixth order WKB method with $\tilde{m}=4$, where $\tilde{m}$ is defined in \cite{Matyjasek:2017psv,Konoplya:2019hlu}, because this choice provides the best accuracy in the Schwarzschild limit and there is hope that this will be the case for more general metrics.

\subsection{Frobenius method}
In order to find accurate values of quasinormal modes we use the method proposed by Leaver~\cite{Leaver:1985ax}. The wave-like equation~(\ref{eq:Q_master}) always has a regular singularity at the horizon $r=r_g$ and the irregular singularity at spatial infinity $r=\infty$. We introduce the new function,
\begin{equation}\label{reg}
\Psi(r)= P (r, \omega) \left(1-\frac{r_g}{r}\right)^{-\imo\omega/F'(r_g)}y(r),
\end{equation}
where the factor $P$ is chosen in such a way that that $y(r)$ is regular for $r_g\leq r<\infty$, once $\Psi(r)$ corresponds to the purely outgoing wave at spatial infinity and the purely ingoing wave at the event horizon. Therefore, we are able to represent $y(r)$ in terms of the Frobenius series:
\begin{equation}\label{Frobenius}
y(r)=\sum_{k=0}^{\infty}a_k\left(1-\frac{r_g}{r}\right)^k.
\end{equation}
Then, using Guassian eliminations in the recurrence relation for the coefficients of the expansion we reduce the problem to solution of an algebraic equation. In addition for quicker convergence we use the Nollert improvement \cite{Nollert:1993zz} in its general form for the n-term recurrence relation suggested in \cite{Zhidenko:2006rs}.

\subsection{Time-domain integration}
In order to find quasinormal modes and, foremost, analyze possible echo-like phenomena we will use the time-domain integration method. We will integrate the wavelike equation in terms of the null-cone variables $$u=\tilde{t}-r_*$$ and $$v=\tilde{t}+r_*,$$
which is, strictly speaking, a graviton cone in the EFT approach \cite{Mukohyama:2022enj} and the "effective" time $\tilde{t}$ is the result of the coordinate transformations, which includes the original spacial coordinate as well \cite{Mukohyama:2022enj}.

Then, we apply the discretization scheme of Gundlach-Price-Pullin \cite{Gundlach:1993tp},
\begin{equation}\label{Discretization}
\Psi\left(N\right)=\Psi\left(W\right)+\Psi\left(E\right)-\Psi\left(S\right) -\Delta^2V\left(S\right)\frac{\Psi\left(W\right)+\Psi\left(E\right)}{4}+{\cal O}\left(\Delta^4\right)\,,
\end{equation}
where the following notation for the points was used:
$N\equiv\left(u+\Delta,v+\Delta\right)$, $W\equiv\left(u+\Delta,v\right)$, $E\equiv\left(u,v+\Delta\right)$, and $S\equiv\left(u,v\right)$. The Gaussian initial data are imposed on the two null surfaces, $u=u_0$ and $v=v_0$. The dominant quasinormal frequencies can be extracted from the time-domain profiles with the help of the Prony method see, e.g.,~\cite{Konoplya:2011qq}.

\section{Quasinormal modes: outburst of overtones and time-domain evolution}\label{QNMsummary}

As can be seen from the wave-like equation and the form of the function $F(r)$ relating the Schwarzschild-like radial coordinate and the tortoise one,  quasinormal modes for each type (channel) of perturbations is described by a different horizon. Therefore quasinormal modes of the Hayward metric studied, for example, in \cite{Konoplya:2022hll}, will be completely different from those studied within the Effective Field Theory. Here, for convenience, we used the units $r_{g}= 1$. If fixing $\mu$ instead of $r_{g}$, which was the units of  \cite{Mukohyama:2023xyf}, then  we reproduce the results of table II in \cite{Mukohyama:2023xyf} both via the Leaver and 6th order WKB methods.
Here, however, we used 7th order WKB with the Pade approximants which is in better concordance with the accurate Frobenius method (see table I). Notice, that our main method here is the Frobenius method, because it is based on the convergent procedure, while WKB formula converges only asymptotically and does not guarantee convergence in each order.

One can see that while the fundamental mode is changed relatively softly, the first few overtones are changing at a much higher rate. A similar phenomenon takes place for perturbations of test fields in the Hayward metric \cite{Konoplya:2022hll}, because there the metric function deviates from the Schwarzschild limit especially strongly near the event horizon.

Using the first order WKB formula and expanding in terms of small $\sigma$ and $1/\ell$ we can find quasinormal modes in analytic form in the eikonal regime $\ell \gg 1$. For this the location of the maximum of the effective potential can be found as follows:
\begin{equation}
r_{max} \approx \frac{3}{2}+ \frac{1}{2 \ell^2}-\frac{\sigma^3}{6} + \frac{7 \sigma^3}{18 \ell^2}.
\end{equation}
Using the above expression in the first order WKB formula we find
\begin{equation}
\omega_n = \ell  \left(\frac{2}{3 \sqrt{3}}+\frac{22 \sigma
   ^3}{81 \sqrt{3}}\right)+\left(\frac{(1-i) - 2 i n}{3
   \sqrt{3}}+\frac{(11- 35 i -70 i n) \sigma ^3}{81
   \sqrt{3}}\right)+O\left(\sigma^6, \frac{1}{\ell
   }\right)
\end{equation}
When $\sigma =0$, this formula goes over into the one for the Schwarzschild black hole (see, for example, eq. (4.5)  \cite{Konoplya:2011qq,Zhidenko:2003wq,Ferrari:1984zz} and references therein). It is worth noticing that the above case of eikonal regime of quasinormal modes, as well as other examples \cite{Konoplya:2017wot,Konoplya:2022gjp,Konoplya:2019hml}, breaks the correspondence between the eikonal quasinormal modes and null geodesics claimed in \cite{Cardoso:2008bp}.

Using the time-domain integration, we can see the evolution of $\Psi$ at a given spacial point as a function of time (see fig. \ref{fig3}). The power-law tails at late times are indistinguishable from the Schwarzschild ones:
\begin{equation}
| \Psi | \sim \tilde{t}^{- (2 \ell +3)}, \quad t \rightarrow \infty.
\end{equation}
Notice, that here the time $ \tilde{t}$ is the result of the transformation  $t \rightarrow \tilde{t}(r,t)$ in the EFT \cite{Mukohyama:2022enj}.

\section{Scattering problem and grey-body factors}\label{Scattering}
\begin{figure}
\centerline{\resizebox{\linewidth}{!}{\includegraphics*{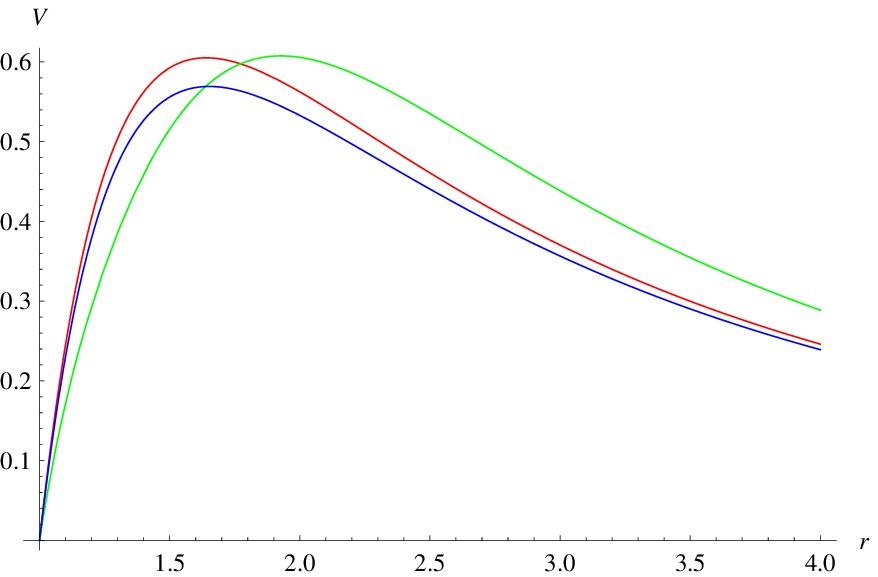}\includegraphics*{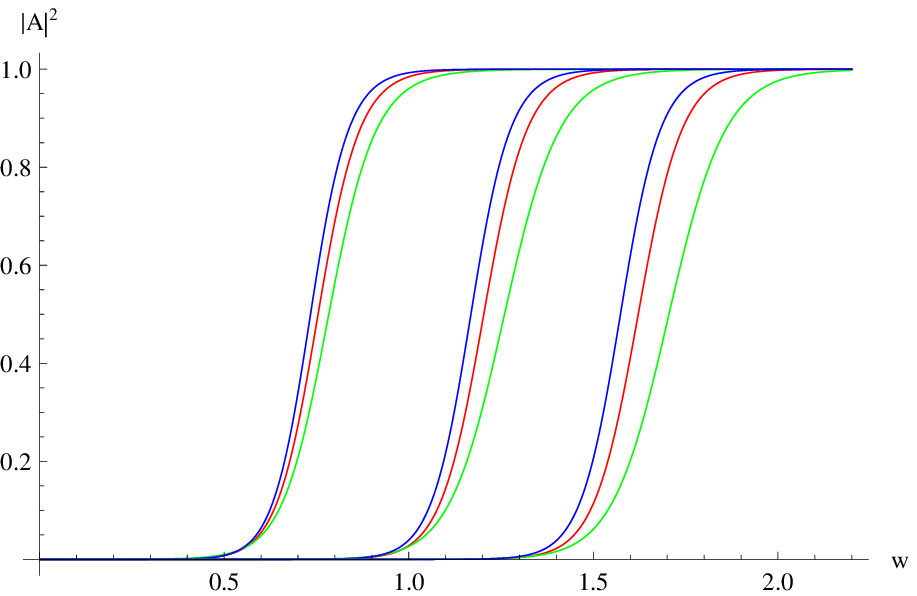}}}
\caption{Left panel: Effective potential for $\ell=2$ perturbations; $\sigma=0$ (red), $\sigma =-0.4$ (blue) and $\sigma =1$ (green). Right panel: grey-body factors for $\ell=2, 3, 4$ (from left to right on the plot); $\sigma=0$ (red), $\sigma =-0.4$ (blue) and $\sigma =0.75$ (green). }\label{fig2}
\end{figure}

The computation of grey-body factors is crucial for determining the proportion of the initial quantum radiation that is reflected back to the event horizon by the potential barrier. Despite the temperature is usually a dominant factor for the intensity and amount of Hawking radiation, in some cases grey-body factors can be more influential than temperature  \cite{Konoplya:2019ppy}.

We will examine the wave equation (\ref{eq:Q_master}) under boundary conditions that allow for incoming waves from infinity. Due to the symmetry of scattering properties, this is equivalent to the scattering of a wave originating from the near-horizon zone. The boundary conditions for scattering in (\ref{eq:Q_master}) are:
\begin{equation}\label{BC}
\begin{array}{ccll}
    \Psi &=& e^{-i\omega r_*} + R e^{i\omega r_*},& r_* \rightarrow +\infty, \\
    \Psi &=& T e^{-i\omega r_*},& r_* \rightarrow -\infty, \\
\end{array}%
\end{equation}
where $R$ and $T$ are the reflection and transmission coefficients.
\par
The effective potential has the form of a potential barrier that decreases monotonically towards both infinities, allowing, thereby, for the application of the WKB approach \cite{Schutz:1985km,Iyer:1986np,Konoplya:2003ii} to determine $R$ and $T$. As $\omega^2$ is real in the scattering problem, the first-order WKB values for $R$ and $T$ will be real \cite{Schutz:1985km,Iyer:1986np,Konoplya:2003ii}, and
\begin{equation}\label{1}
\left|T\right|^2 + \left|R\right|^2 = 1.
\end{equation}
Once the reflection coefficient is obtained, one can find the transmission coefficient for each $\ell$
\begin{equation}
\left|{\pazocal
A}_{\ell}\right|^2=1-\left|R_{\ell}\right|^2=\left|T_{\ell}\right|^2.
\end{equation}

In order to study the reflection and transmission coefficients we will use the higher order WKB formula \cite{Konoplya:2003ii}. However, this formula is not suitable for very small values of $\omega$, which correspond to almost complete wave reflection and have negligible contributions to the overall energy emission rate. For this regime, we employed extrapolation of the WKB results at a given order to smaller $\omega$. According to \cite{Schutz:1985km,Iyer:1986np}, the reflection coefficient can be expressed as follows,
\begin{equation}\label{moderate-omega-wkb}
R = (1 + e^{- 2 i \pi K})^{-\frac{1}{2}},
\end{equation}
where $K$ is determined by solving the equation
\begin{equation}
K - i \frac{(\omega^2 - V_{max})}{\sqrt{-2 V_{max}^{\prime \prime}}} - \sum_{i=2}^{i=6} \Lambda_{i}(K) =0,
\end{equation}
involving the maximum effective potential $V_{max}$, the second derivative $V_{max}^{\prime \prime}$ with respect to the tortoise coordinate, and higher order WKB corrections $\Lambda_i$.

The WKB formula for finding grey-body factors is known to provide reasonable accuracy for further estimation of the intensity of Hawking radiation and was, therefore, used in a number of papers (see, for example, \cite{Konoplya:2020cbv,Konoplya:2020jgt}).
From fig. \ref{fig2} one can see that the larger is $\sigma$, the smaller is the grey-body factors, that is the bigger portion of particles is reflected by the effective potential. This can easily be understood from the behavior of the effective potential which becomes higher (i.e. more difficult to penetrate) for larger $\sigma$.

\section{Conclusions}

Scalar-tensor theories have been recently actively studied as alternatives to the Einstein gravity  (see, for example, \cite{Mukohyama:2022enj,Joshi:2021azw} and references therein). Here we considered quasinormal modes and grey-body factors of gravitational perturbations of the Hayward black hole with a scalar hair built within the Effective Field Theory. The initial study of $\ell =2$, $n=0$ frequency has been recently suggested in \cite{Mukohyama:2023xyf}. Here we studied quasinormal frequencies for various $n$ and $\ell$ in more details. Our main finding here is that the first few overtones deviate from their Schwarzschild limit at a much higher rate than the fundamental mode and this happens because of the deformation of the spacetime near the horizon. The integration of the wave-like equation in time-domain confirms the dominant quasinormal frequencies found by other methods and shows that the power-law tails remain the same as in the Einstein theory.  In the eikonal regime of large multipole numbers $\ell$ the analytical formula for quasinormal modes has been obtained. In addition we have found grey-body factors for the above gravitational perturbations and showed that the bigger is $\sigma$ (at a fixed radius of the horizon $r_g$), the smaller is the grey-body factors, because the effective potential becomes higher in this case.

Our work could be extended in a number of ways. Once the corresponding wave-like equations were obtained for the test fields, the corresponding quasinormal modes and grey-boday factors could be obtained in a similar way. If, in addition, the temperature is properly defined \cite{Mukohyama:2009um,Mukohyama:2009rk}, one could use the above constituents to estimate the intensity of Hawking radiation.

\begin{acknowledgments}
The author acknowledges A. Zhidenko and S. Mukohyama for useful discussions.
\end{acknowledgments}

\end{document}